\documentstyle[twoside,fleqn,epsfig,latexsym,amssymb,espcrc3,graphicx,graphics]{article}
\def\ct#1{{\cal #1}}

\def\bc{\begin{center}}
\def\ec{\end{center}}
\def\bs{\begin{slide}}
\def\es{\end{slide}}

\newcommand{\gev}{\mathrm{GeV}}

\newcommand{\ba}{\begin{array}}
\newcommand{\ea}{\end{array}}
\newcommand{\bes}{\begin{equation*}}
\newcommand{\ees}{\end{equation*}}
\newcommand{\beqns}{\begin{eqnarray*}}
\newcommand{\eeqns}{\end{eqnarray*}}

\def\vdir{v\kern-8.75pt\raise0.15ex\hbox{${\scriptstyle /}$}}
\def\pdir{p\kern-7.8pt\raise0.2ex\hbox{\Big{/}}}
\def\ddir{D\kern-13.75pt\raise0.15ex\hbox{\Big{/}}}
\def\partdir{\partial\kern-7.6pt\raise0.25ex\hbox{/}}
\def\ddirp{D_{\kern-5pt\perp}\kern-22pt\raise0.15ex\hbox{\Big{/}}\kern+5.5pt}

\def\ct#1{{\cal #1}}

\def\slash#1{\setbox0=\hbox{$#1$}\dimen0=\wd0 \setbox1=\hbox{/} \dimen1=\wd1 \ifdim\dimen0>\dimen1 \rlap{\hbox to \dimen0{\hfil/\hfil}} #1 \else \rlap{\hbox to \dimen1{\hfil$#1$\hfil}} / \fi}                                         

\def\gev{\mbox{ GeV}}

\def\op{O}

\def\lrvec#1{\setbox0=\hbox{$#1$}
    \setbox1=\hbox{$\scriptstyle\leftrightarrow$}
    #1\kern-\wd0\smash{\raise\ht0\hbox{$\raise1pt\hbox{
$\scriptstyle\leftrightarrow$}$}}\kern-\wd1\kern\wd0}

\def\spose#1{\hbox to 0pt{#1\hss}}
\def\ltapprox{\mathrel{\spose{\lower 3pt\hbox{$\mathchar"218$}}
 \raise 2.0pt\hbox{$\mathchar"13C$}}}
\def\gtapprox{\mathrel{\spose{\lower 3pt\hbox{$\mathchar"218$}}
 \raise 2.0pt\hbox{$\mathchar"13E$}}}
\def\inapprox{\mathrel{\spose{\lower 3pt\hbox{$\mathchar"218$}}
 \raise 2.0pt\hbox{$\mathchar"232$}}}

\newcommand{\nn}{\nonumber}

\newcommand{\<}{\langle}
\renewcommand{\>}{\rangle}
\newcommand{\la}{\langle}
\newcommand{\ra}{\rangle}

\newcommand{\msbar}{\overline{\mbox{\scriptsize MS}}}

\newcommand{\be}{\begin{equation}}
\newcommand{\ee}{\end{equation}}
\newcommand{\bi}{\begin{itemize}}
\newcommand{\ei}{\end{itemize}}
\newcommand{\bea}{\begin{eqnarray}}
\newcommand{\eea}{\end{eqnarray}}

\newcommand{\rar}{\rightarrow}

\newcommand{\beq}{\begin{equation}}
\newcommand{\eeq}{\end{equation}}
\newcommand{\beqn}{\begin{eqnarray}}
\newcommand{\eeqn}{\end{eqnarray}}

\newcommand{\vettq}{{\bf q}}
\newcommand{\vettp}{{\bf p}}

\newcommand{\vettx}{{\bf x}}

\newcommand{\ditre}{$\Delta I=3/2$ }

\title{\vspace*{-0.0cm}Kaon Weak Matrix Elements with Wilson Fermions\thanks{Talk presented by Mauro Papinutto.}}

\author{$\mathrm{SPQ_{CD}R}$ Collaboration:~D.~Becirevic\address{\vspace*{-0.29cm}Dip. di Fisica, 
Universit\`a ``La Sapienza" and INFN-Roma I, P.le A. Moro 2, I-00185 Rome,
Italy},
~Ph.~Boucaud\address{\vspace*{-0.29cm}Laboratoire de Physique Th\'eorique (b\^at.210),
Universit\'e de Paris XI, 91405 Orsay Cedex, France}, 
V.~Gim\'enez\address{\vspace*{-0.29cm}Dep.~de F\'{\i}s.Te\`orica and IFIC, Univ.~de Val\`encia, Dr.~Moliner 50, E-46100, Burjassot, Val\`encia,
Spain}, 
C.-J.~D.~Lin\address{\vspace*{-0.29cm}Department of Physics and Astronomy, University of
Southampton, Southampton SO17 1BJ, England},
V.~Lubicz\address{\vspace*{-0.29cm}Dip. di Fisica, Univ. di Roma Tre and INFN - Roma III, 
Via della V. Navale 84, I-00146 Rome, Italy},   
G.~Martinelli$^{\rm a}$,
M.~Papinutto\address{\vspace*{-0.0cm}DESY, Theory Group, Notkestrasse
 85, D-22607 Hamburg, Germany},
C.~T.~Sachrajda$^{\rm d}$}
\begin{document}

\begin{abstract}
\vspace*{-0.2cm}
We present results of several numerical studies with
Wilson fermions relevant for kaon physics. We compute the $B_K$
parameter by using two different methods and extrapolate to the continuum
limit. Our preliminary result is $B^{\msbar}_K(2\gev)=0.66(7)$.  
$\Delta I=3/2$ $K\rar \pi \pi$ matrix elements (MEs) are obtained by 
using the next-to-leading order (NLO) expressions derived in 
chiral perturbation theory (ChPT) in which the low energy 
constants (LECs) are determined by the lattice results computed at 
unphysical kinematics. From the simulation at
$\beta\!\!=\!\!6.0$ our (preliminary) results read: 
$\la\pi\pi|\op_{7}^{\msbar}(2 \gev)|K\ra_{I=2}\!\!=\!\!0.14(1)(1) \gev^3$
and $\la\pi\pi|\op_{8}^{\msbar}(2 \gev)|K\ra_{I=2}\!\!=\!\!0.69(6)(6) \gev^3$.

\vspace*{-0.4cm}
\end{abstract}
 
\maketitle

\section{$K^0$--$\bar K^0$ MIXING}

\vspace*{-0.2cm}
The main problem of the computation of the $K^0$--$\bar K^0$
mixing amplitudes with Wilson fermions is the lattice operator mixing (not
present in the continuum) introduced by the explicit breaking of the 
chiral symmetry due to the Wilson term. In order to compute the
amplitudes one has to subtract contributions of operators with 
different na\"{\i}ve chiralities. Two alternative proposals which avoid the
operator subtractions have been recently suggested: the first uses twisted
mass QCD~\cite{pena} while the second uses suitable Ward Identities 
(WIs)~\cite{bkws}. 
A first comparison of the standard method (with subtractions) to the WIs
method (without subtractions) has been presented in ref.~\cite{damir}
for the case of the parameter $B_K$, defined as

\vspace*{-0.2cm}\noindent\[
\qquad\langle \bar K^0 \vert O_1
\vert K^0 \rangle \equiv \frac{8}{3} f_{K}^{2} m_{K}^{2}
B_{K},
\]

\vspace*{-0.2cm}\noindent with $O_1\!=\!Q_1\!-\!\ct{Q}_1\!=\!{\bar s}^a 
\gamma_{\mu} (1-\gamma_5) d^a{\bar s}^b \gamma_{\mu}   (1-\gamma_5)
d^b$. $Q_1$ is
the parity conserving part of $O_1$ and under renormalization mixes 
with operators of different na\"{\i}ve chiralities, while $\ct{Q}_1$ is its parity
violating part and due to $\ct{CPS}$ symmetry is multiplicatively
renormalizable. The basic idea for the WIs method is based on 
the following WI obtained from a $\tau_3$ axial rotation:\\

\vspace*{-0.7cm}\noindent\scalebox{.95}{\parbox{18cm}{
\bea m \sum^{}_{z} \langle \Pi^{0}(z) P(t_x,\vettq)  
\ct{Q}_{1}(0) P(t_y,\vettq)  \rangle + {\cal O}(a) = \nn\\
\langle P(t_x,\vettq) Q_{1}(0) P(t_y,\vettp) \rangle +
\textrm{(rot. sources),} \nn
\eea}}

\vspace*{-0.1cm}\noindent 
where $P(t,\vettp)= \sum_{\vettx} e^{-i\vettp\vettx} \bar
d(t,\vettx)\gamma_{5}s(t,\vettx)$
and $\Pi^{0}= \bar d \gamma_{5}d-\bar u\gamma_{5}u$. $SU(3)$ symmetry
together with charge conjugation $\ct{C}$ imply that the terms involving
the rotation of the sources are identically zero. Instead of
computing the three point function involving $Q_{1}$ one
can compute the four point function involving $\ct{Q}_{1}$, thus
avoiding lattice operator mixing. From now on the procedure to obtain $B_K$ 
is common to both methods and can be found in ref.~\cite{firstbk}. Note,
however, that ${\cal O}(a)$ effects present in both methods may be different.
Our simulation has been performed at three values of $\beta$, 
by keeping roughly the same physical volume. 

\vspace*{0.2cm}\scalebox{.84}{\parbox{1cm}
{\begin{tabular}{cccc}\hline\hline
$\beta$ & 6.0 & 6.2 & 6.4 \\
\hline 
volume & $16^3\times 52$ & $24^3\times 64$ & $32^3\times 70$\\ 
$\#$ confs. & 500 & 200 & 135\\ 
$a^{-1}$ from $K/K^*$ & 2.05(7) & 2.68(11) & 3.44(10)\\ 
$a M_P$ & 0.30-0.39 & 0.20-0.35 & 0.15-0.23\\ 
\hline \hline
\end{tabular}}}

\begin{figure}[!h]
\vspace*{-1.0cm}
\begin{center}
\epsfig{figure=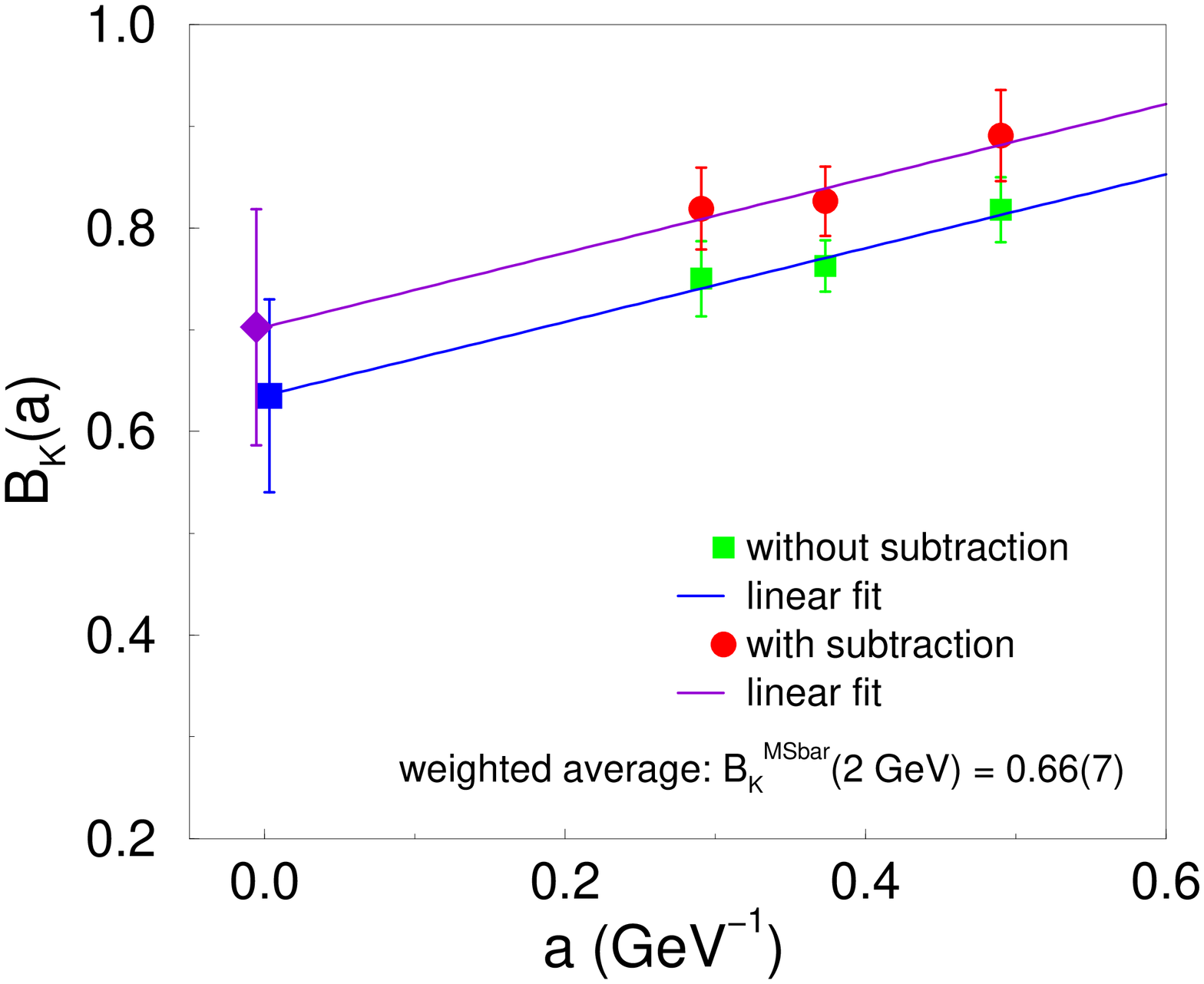,angle=270,width=0.57\linewidth}
\vspace*{-1.1cm}
\caption{\sl Continuum limit for $B^{\msbar}_K(2\gev)$}
\vspace*{-1.4cm}
\label{continuum}
\end{center}
\end{figure}

\vspace*{0.1cm}\noindent
The values of $B^{\msbar}_K(2\gev)$ for each value of $\beta$ are given
in Table~1, where the last column contains the values obtained 
from a linear extrapolation to the continuum limit (see 
Fig.~\ref{continuum}). The average of the two
gives $B^{\msbar}_K(2\gev)=0.66(7)$. We are currently preparing a
simulation at a larger $\beta$ to discard, eventually, the point 
at $\beta=6.0$ which could lie outside the region linear in $a$.  
\begin{table}[!t]
\vspace*{-0.1cm}\scalebox{.81}{\parbox{9cm}
{\begin{tabular}{ccccc}\hline\hline
$\beta$ & 6.0 & 6.2 & 6.4 & $\infty$\\
\hline 
$a$ (fm) & 0.096(3) & 0.073(3) & 0.057(2) & 0\\ 
no subt. & 0.818(32) & 0.763(25) & 0.750(37) & 0.634(95)\\ 
stand. & 0.890(45) & 0.826(34) & 0.819(40) & 0.70(12)\\
\hline \hline
\end{tabular}}}

\vspace*{0.15cm}Table 1. {\sl Values of $B^{\msbar}_K(2\gev)$ in
function of $a$}
   
\vspace*{-0.7cm}
\end{table}

\section{$K\rar\pi$ MEs IN THE CHIRAL LIMIT}
\label{ktopi}

The $K\!\!\rar\!\!\pi$ MEs of $O^{3/2}_7$,$O^{3/2}_8$ 
(defined in ref.~\cite{giusti})
can be related, in the chiral limit, to those of the left-right
operators of the $\Delta S\!\!=\!\!2$ basis as explained in
ref.~\cite{giusti}. By using the soft
pion theorem $\langle\pi\pi| O_{7,8}|K^0\rangle_{I=2}=-i/(\sqrt{2}f_\pi)
\langle\pi^{+}|O^{3/2}_{7,8}|K^{+}\rangle$, 
we obtain the $\Delta I\!\!=\!\!3/2$ MEs of the electro-weak 
penguin (EWP) operators $O_7$,$O_8$ (defined in Eq.~(1) of ref.~\cite{lin1}) 
in the chiral limit. Our results (in the $\msbar$ scheme at $2 \gev$) read:    

\vspace*{0.2cm}\noindent\scalebox{.9}
{\parbox{9cm}{\begin{tabular}{cccc}\hline\hline
$\beta$ &\hspace*{-0.3cm} 6.0 &\hspace*{-0.3cm} 6.2 &\hspace*{-0.3cm} 6.4 \\
\hline 
$\!\!\!\!\<\pi\pi| O_7|K^0\>_{I=2}[\!\!\gev^3\!]$ &\hspace*{-0.3cm}
0.132(10) &\hspace*{-0.3cm} 0.136(14) &\hspace*{-0.3cm} 0.136(10) \\ 
$\!\!\!\!\<\pi\pi| O_8|K^0\>_{I=2}[\!\!\gev^3\!]$ &\hspace*{-0.3cm}
0.658(44) &\hspace*{-0.3cm} 0.604(44) & \hspace*{-0.3cm} 0.533(34) \\
\hline \hline
\end{tabular}}}

\vspace*{0.2cm}\noindent For $O_8$ discretization errors seem quite
large and we are currently investigating this feature.

\section{\ditre $K\rar\pi\pi$ AMPLITUDES}

The procedure to extract these amplitudes by using ChPT
at NLO has been explained in ref.~\cite{lin1,mauro,lin2}. 
We are interested in $K\rar\pi\pi$ matrix elements of the EWPs and
of the operator $O_4$ (defined in Eq.~(1) of ref.~\cite{lin1}) which
mainly determines the physical $K^+\rar\pi^+\pi^0$ decay amplitude. 
In practice we
work with an unphysical kinematics (called SPQR) where the kaon is at
rest, one pion is at rest and the other has either momentum 0 or
momentum $2\pi/L$. We have 3 pion masses and 3 kaon masses for each pion
one ($M_\pi \in [0.51,0.88]\gev,\, M_K\in[0.51,1.46]\gev$), and the energy 
is in general not conserved (it 
can be injected through the weak operator~\cite{LMST}). We fit 
the numerical data to the expressions obtained at
NLO in quenched ChPT~\cite{lin1,lin2}. For example, for the EWPs we have 
\vspace*{1.0cm}\hspace*{-0.2cm}\scalebox{.81}{\parbox{18cm}{\bea 
  &  {\cal M}_{7,8}^{\textrm{\tiny{SPQR}}}\!\!\!\! &= 
\gamma^{7,8}\left(1+\frac{M_K^2}{(4\pi f)^2}
(\textrm{chiral logs})^{\textrm{\tiny{SPQR}}}
\right)\nn
\\ &&- \left(\delta^{7,8}_3 + \frac{\delta^{7,8}_2}{2} - 
\frac{\delta^{7,8}_1}{2}\right)\, M_K \,( M_\pi+E_\pi) \nonumber
 \\&&+ \left(  \delta^{7,8}_4  + \delta^{7,8}_5
\right) \left( 2  M^2_K + 4 M^2_\pi \right)
 \nonumber \\ &&+ \delta^{7,8}_6 \left( 4  M^2_K + 2 M^2_\pi \right)-
(\delta^{7,8}_{1}+\delta^{7,8}_{2})\, E_\pi\, M_\pi,\;\qquad\quad(1) 
\nonumber 
\eea}}
\begin{figure}[!t]
\vspace*{-0.8cm}
\begin{center}
\epsfig{figure=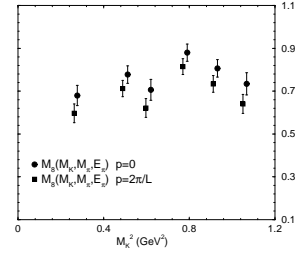,angle=270,width=0.51\linewidth}
\vspace*{-0.9cm}
\caption{\sl Numerical data for the ME of $O_8$}
\vspace*{-1.2cm}
\label{O8proceed}
\end{center}
\end{figure}     

\vspace*{-1.0cm}\noindent where $\gamma^{7,8}$ is the LEC of the 
LO while the $\delta^{7,8}_i$ are the LECs of the NLO. For the moment 
we have not considered the finite
volume correction (the Lellouch-L\"uscher (LL) factor~\cite{LL}) which is not
yet known for the SPQR kinematics. In this way we determine some combinations
of the LECs of the NLO and we use them to compute the values of the
physical amplitudes (see refs.~\cite{lin1,lin2} and refs. therein). 
Again in the case of the EWPs we have 

\vspace*{-0.1cm}\noindent\hspace*{-0.0cm}\scalebox{.81}{\parbox{18cm}{\bea
&{\cal M}^{\tiny{\textrm{phys}}}_{7,8} \!\!\!\!&=
\gamma^{7,8}\left(1+\frac{m_K^2}
{(4\pi f)^2}(\textrm{chiral logs})^{\textrm{\tiny{phys}}}\right)\nn \\
&&+ \left( -\delta^{7,8}_{2}-\delta^{7,8}_{3} + 2 \delta^{7,8}_{4} + 2 
\delta^{7,8}_{5} + 4\delta^{7,8}_{6}\right) \, m^{2}_{K}\nn\\
 &&+ \left(  \delta^{7,8}_{1}+ \delta^{7,8}_{2}+ 4 \delta^{7,8}_{4} + 
4 \delta^{7,8}_{5} + 2\delta^{7,8}_{6 }\right)\,  m^{2}_{\pi}.\nonumber \eea}}

\vspace*{-0.0cm}\noindent For the EWPs, 6 counterterms of the NLO 
contribute in the SPQR kinematics and, as shown in Eq.~(1),
we can determine only 4 
combinations of the corresponding LECs. These are enough to compute 
the physical amplitude. For $O_4$ there are again 6 counterterms at 
the NLO but all of the corresponding LECs have to be separately determined. 
Consequently, in the second case the fit is much more complicate. 

For the EWPs, the analysis at LO (which is ${\cal O}(p^0)$ and thus 
correspond to fit the numerical data to a constant) gives the results

\vspace*{0.25cm}\noindent\scalebox{0.88}{\parbox{18cm}{
\begin{tabular}{cccc}
\hline\hline
$M_K,E_\pi <$ & 1.5 \gev & 1.0 \gev & 0.8 \gev\\ \hline
$\<\pi\pi| O_7|K^0\>_{I=2}$ & 0.076(6)& 0.080(6)& 0.094(6)\\
$\<\pi\pi| O_8|K^0\>_{I=2}$ &0.909(45) & 0.860(45) & 0.803(46)\\
\hline
\hline
\end{tabular}}}

\begin{figure}[!t]
\vspace*{-0.5cm}
\begin{center}
\mbox{\hspace*{-3.5cm}
\epsfig{figure=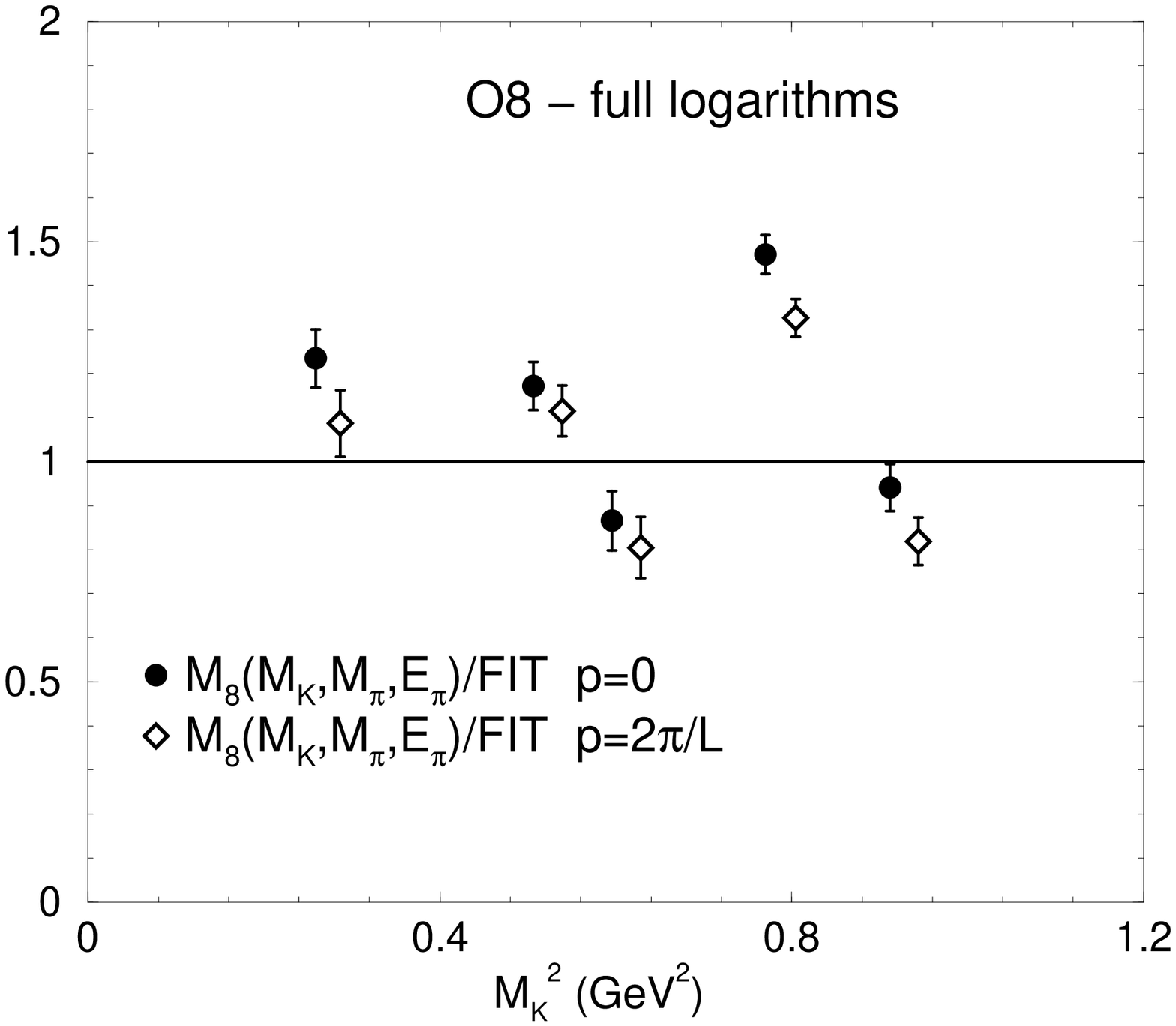,angle=270,width=0.50\linewidth}
\put(0,0){\epsfig{figure=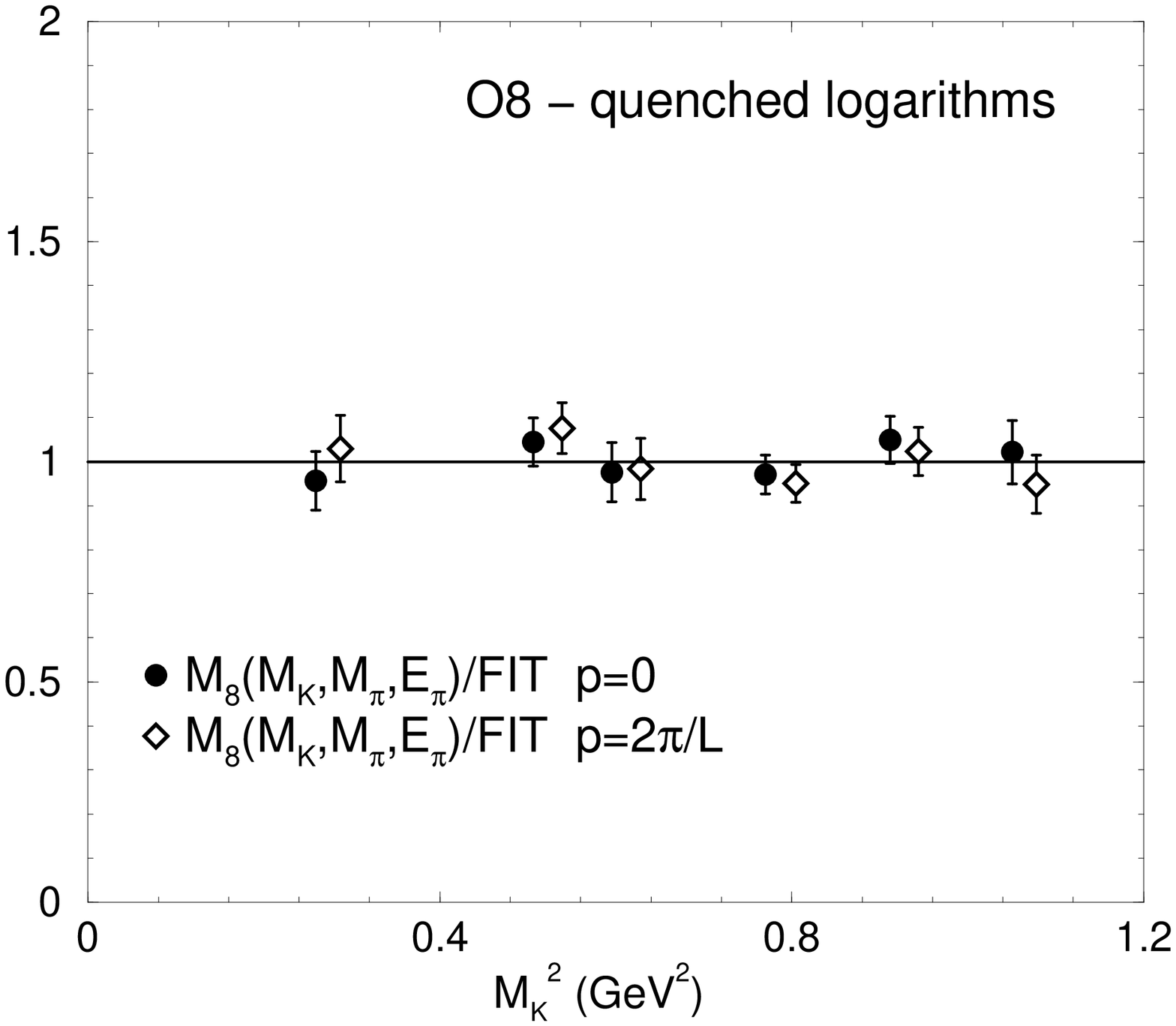,angle=270,width=0.50\linewidth}}}
\vspace*{-0.9cm}
\caption{\sl Fits to ${\cal M}_8$ with full and quenched logs}
\vspace*{-1.0cm}
\label{O8fit}
\end{center}
\end{figure}

\begin{figure}[!t]
\vspace*{-0.1cm}\hspace*{1.2cm}
\epsfig{figure=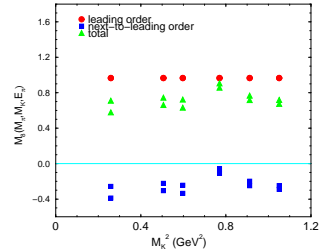,angle=270,width=0.54\linewidth}
\vspace*{-0.95cm}
\caption{\sl Various contributions to the fit of ${\cal M}_8$}
\vspace*{-0.8cm}
\label{various}
\end{figure}

\vspace*{0.25cm}\noindent where we varied the number of data points used
in the fit by choosing only points with masses and
energies below a certain cut-off. The dependence of the results on the
cut-off is due to the fact that higher orders in ChPT are not negligible
(as one can argue from Fig.~\ref{O8proceed} in the case of $O_8$).    
We thus introduce the NLO ${\cal O}(p^2)$ contribution in the
fit. In Fig.~\ref{O8fit} we show the quality of the fits obtained by using
either the logarithms of full ChPT or those of quenched ChPT. 
We see that, according to what expected~\cite{lin3}, only
the quenched logarithms have the appropriate form to fit the data. On
the other hand, the fit is not very sensitive to the quenching parameters
$\delta$ and $\alpha$~\cite{bernard} and, in the lacking of their precise 
determination, we set their values to be
$\delta=0.1$ and $\alpha=0.0$. The contributions of 
the LO and NLO to the fit are shown in Fig.~\ref{various}. The fit now
shows much less dependence on the cut-off and we quote the results 
obtained with $M_{K}, E_{\pi} < 0.8$ GeV
as our central values, and use those with other cut-offs to estimate
the systematic error due to the higher-order corrections in the 
chiral expansion.  This leads to the preliminary results
(in which the finite-volume LL corrections have not been included): 

\vspace*{-0.1cm}\noindent\bea\la\pi\pi|\op_{7}^{\msbar}(2 \gev)|K\ra_{I=2} = 0.14(1)(1) \gev^3,&&\nn\\
\la\pi\pi|\op_{8}^{\msbar}(2 \gev)|K\ra_{I=2}=0.69(6)(6)
\gev^3.&\;(2)&\nn\eea 

\vspace*{-0.1cm}Note, however, that we can not claim to be 
sensitive to the chiral
logarithms. In fact we could have fitted our numerical data by including
only the counterterms of the NLO and not the logarithms. Also in this case we
would have obtained a very good fit (of quality comparable to that which
includes the chiral logarithms) and results which depend very weakly 
on the cut-off and which are very close to 
those obtained by including the logarithms: 
$\la\op_{7}\ra_{I=2}\!\!=\!\!0.13(1)(1) \gev^3$
and $\la\op_{8}\ra_{I=2}\!\!=\!\!0.64(6)(2) \gev^3$.    
We also note that this determination and the one in
Eq.~(2) are both in reasonable agreement
with the results obtained from $K\!\!\!\rar\!\!\pi$ MEs in
Sec.\ref{ktopi} (at $\beta\!\!=\!\!6.0$). Finally, it is important 
to remark that the analysis at LO would lead to a large systematic 
error (as large as $40\%$ in the case of $O_7$).   

Also in the case of $O_4$ the NLO contribution turns out to be large. 
The fit is more problematic due to the large number of parameters 
and the errors on some of the NLO
LECs (and therefore on the whole NLO contribution) is quite 
large. As an indicative result (obtained with $M_K,E_\pi< 1\gev$)
we quote ${\cal A}(K^+\!\!\!\rar\!\!\!\pi^+\pi^0)\!\!\!=\!\!\!0.0097(18)^{LO}
\!\!\!+\!\!\!0.0039(85)^{NLO} \gev^3$ (the experimental value is 
${\cal A}^{\tiny{\textrm{exp}}}\!\!\!=\!\!\!0.0104\gev^3$). 
We are presently trying to consider other
kinematical configurations which could improve the quality of the fit.


\end{document}